\documentclass[prb,superscriptaddress,twocolumn,showpacs,preprintnumbers,amsmath,amssymb]{revtex4}
\usepackage{graphicx}% Include figure files
\usepackage{dcolumn}% Align table columns on decimal point
\usepackage{bm}% bold math
\usepackage{color}

\newcommand{\bea}{\begin{eqnarray}}
\newcommand{\eea}{\end{eqnarray}}

\newcommand{\be}{\begin{equation}}
\newcommand{\ee}{\end{equation}}

\newcommand{\K}{\mathbf{k}}

\newcommand{\R}{\mathbf{r}}

\newcommand{\si}{\mathbf{S}_i}

\newcommand{\sj}{\mathbf{S}_j}

\newcommand{\Sa}{\mathbf{S}_A}
\newcommand{\Sb}{\mathbf{S}_B}
\newcommand{\Sc}{\mathbf{S}_C}

\newcommand{\Saz}{S_A^z}
\newcommand{\Sbz}{S_B^z}
\newcommand{\Scz}{S_C^z}

\newcommand\kagome{kagome}
\newcommand\sat{{\rm sat}}

\begin{document}

\title{Magnetization Process of the Classical Heisenberg Model\\ on the Shastry-Sutherland Lattice}

\author{M.\ Moliner}
\email{moliner@lpt1.u-strasbg.fr}
\affiliation{Laboratoire de Physique Th\'eorique, Universit\'e Louis Pasteur, UMR 7085 CNRS, 67084 Strasbourg, France}

\author{D.\ C.\ Cabra}
\affiliation{Laboratoire de Physique Th\'eorique, Universit\'e Louis Pasteur, UMR 7085 CNRS, 67084 Strasbourg, France} 
\affiliation{Departamento de F\'{\i}sica, Universidad Nacional de la Plata, C.C.\ 67, (1900) La Plata, Argentina}
\affiliation{Facultad de Ingenier\'\i a, Universidad Nacional de Lomas de Zamora, Cno.\ de Cintura y Juan XXIII, (1832) Lomas de Zamora, Argentina}

\author{A.\ Honecker}
\affiliation{Institut f\"ur Theoretische Physik, Georg-August-Universit\"at G\"ottingen, 37077 G\"ottingen, Germany}

\author{P.\ Pujol}
\affiliation{Laboratoire de Physique Th\'eorique, IRSAMC, Universit\'e Paul Sabatier, CNRS, 31062 Toulouse, France}

\author{F.\ Stauffer}

\affiliation{RBS Service Recherche \& D\'eveloppement, Strasbourg -- Entzheim, 67836 Tanneries Cedex, France}

\date{\today}

%%%%%%%%%%%%%%%%%%%%%%%%%%%%%%%%%%%%%%%%%%%%%%%%%%%%%%%%%%%%%%%%%%%%%%%%%%%%%%%%
%%%%%%%%%%%%%%%%%%%%%%%%%%%%%%%%%%%%%%%%%%%%%%%%%%%%%%%%%%%%%%%%%%%%%%%%%%%%%%%%
\begin{abstract}

We investigate classical Heisenberg spins on the Shastry-Sutherland lattice and under an external magnetic field. A detailed study is carried out both analytically and numerically by means of classical Monte-Carlo simulations. Magnetization pseudo-plateaux are observed around $1/3$ of the saturation magnetization for a range of values of the magnetic couplings. We show that the existence of the pseudo-plateau is due to an entropic selection of a particular collinear state. A phase diagram that shows the domains of existence of those pseudo-plateaux in the $(h, T)$ plane is obtained.
\end{abstract}

%, 05.50.+q}

% Details of the PACS that were used:
% 75.10.-b 	General theory and models of magnetic ordering (see also 05.50.+q Lattice theory and statistics)
% 75.10.Hk 	Classical spin models
% 75.30.Ds 	Spin waves (for spin-wave resonance, see 76.50.+g)
% 75.30.Kz 	Magnetic phase boundaries (including magnetic transitions, metamagnetism, etc.)
% 75.40.Cx 	Static properties (order parameter, static susceptibility, heat capacities, critical exponents, etc.) 
% 75.60.Ej 	Magnetization curves, hysteresis, Barkhausen and related effects
% 05.50.+q	Lattice theory and statistics (Ising, Potts, etc.) (see also 64.60.Cn Order-disorder transformations, and 75.10.Hk Classical spin models)

%\keywords{Shastry-Sutherland lattice, classical spins, magnetization plateaux, triangular lattice}

\maketitle

%%%%%%%%%%%%%%%%%%%%%%%%%%%%%%%%%%%%%%%%%%%%%%%%%%%%%%%%%%%%%%%%%%%%%%%%%%%%%%%%
%%%%%%%%%%%%%%%%%%%%%%%%%%%%%%%%%%%%%%%%%%%%%%%%%%%%%%%%%%%%%%%%%%%%%%%%%%%%%%%%
\section{\label{sec:intro}Introduction}

\begin{figure}
 \vspace{2mm}
 \begin{tabular}{cc}
  \includegraphics[width=0.24\textwidth]{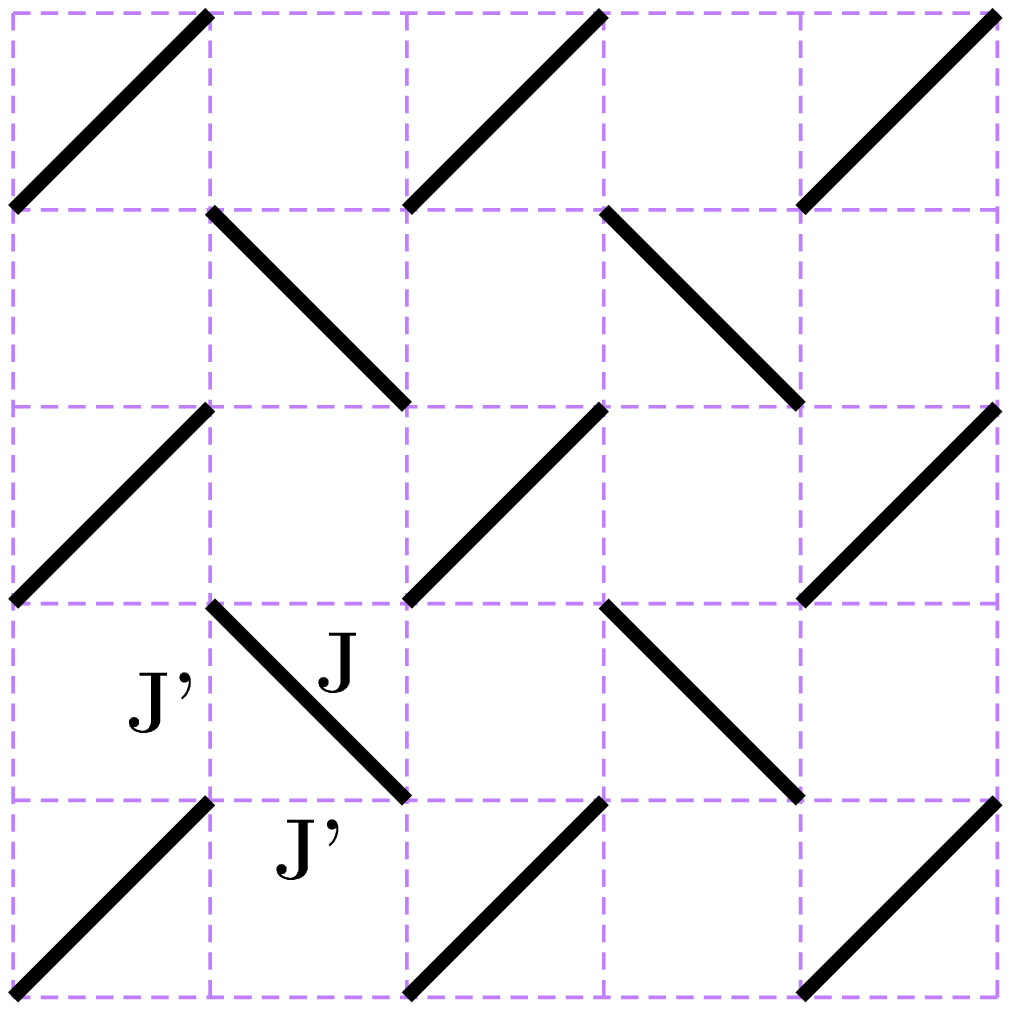} &
  \includegraphics[width=0.23\textwidth]{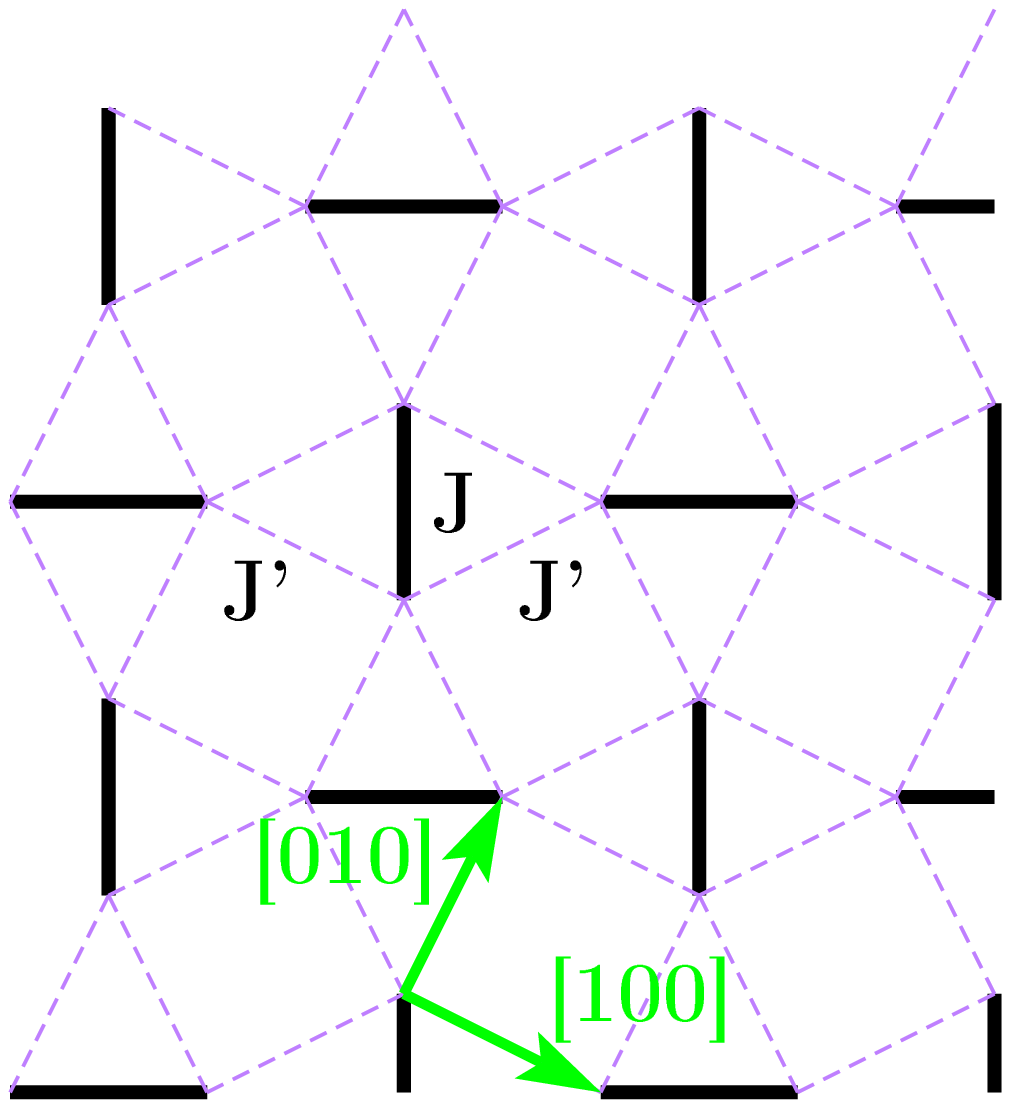}
 \end{tabular}
 \caption{\label{fig:SSL} (color online) The Shastry-Sutherland lattice (left panel) and the topologically identical
  structure realized in the $(001)$ plane of SrCu$_2$(BO$_3$)$_2$ and rare-earth tetraborides (right panel).
  $J'$ bonds are the magnetic couplings along the edges of the squares (clear dashed lines) and $J$ bonds (black bold lines)
  are the diagonal dimer couplings.}
\end{figure}

A spin system is frustrated when all local interactions between spin pairs cannot be satisfied at the same time. Frustration can arise from competing interactions or/and from a particular geometry of the lattice, as seen in the triangular lattice. Frustrated systems have been intensively studied over the last decades and they were found to present very rich behavior such as large ground-state degeneracies. A recent review on frustrated systems is presented in Ref.~\onlinecite{review_frustration}.\\
The Shastry-Sutherland lattice (SSL) was considered more than 20 years ago by Shastry and Sutherland as an interesting example of a frustrated quantum spin system with an exact ground state.\cite{SS} It can be described as a square lattice with $J'$ antiferromagnetic couplings between nearest neighbors and additional $J$ antiferromagnetic couplings between next-nearest neighbors in every second square (see Fig.~\ref{fig:SSL} left panel).\\ 
This lattice attracted much attention after an experimental realization was identified in SrCu$_2$(BO$_3$)$_2$,\cite{SrCu2(BO3)2_dimer_GS_expe} a compound which was first synthesized by Smith and Keszler.\cite{first_expe_SSL_ever} SrCu$_2$(BO$_3$)$_2$ exhibits a layered structure of Cu(BO$_3$) planes separated by magnetically inert Sr atoms. The Cu$^{2+}$ ions carry a spin $S=1/2$ and are located on a lattice which is topologically equivalent to the SSL (see Fig.~\ref{fig:SSL} right panel).
The quantum SSL was widely studied both theoretically and experimentally (see Ref.~\onlinecite{review_SrCu2(BO3)2} for a review). As the ratio $J'/J$ varies, it presents a rich phase diagram with quantum phase transitions between the gapped dimer singlet ground state ($J'/J \lesssim 0.68$) that was originally discussed by Shastry and Sutherland,\cite{SS} a plaquette resonating valence bond state ($J'/J \lesssim 0.86$), and a gapless magnetic state.\cite{SrCu2(BO3)2_3phases,quad_shasu} Theoretical studies on the quantum SSL predicted the existence of plateaux in the magnetization curve for the rational values of the saturated magnetization $M/M_{\sat} = 1/8$, $1/4$, $1/3$ and $1/2$: Momoi and Totsuka \cite{SrCu2(BO3)2_1/3_plateau_th, SrCu2(BO3)2_1/3_and_1/2_plateau_th} showed that the stripe order of the dimer triplets explains the $M/M_{\sat} = 1/2$ and $1/3$ plateaux and the latter was predicted to be the broadest. Experimentally, magnetization plateaux were observed
\cite{SrCu2(BO3)2_dimer_GS_expe, SrCu2(BO3)2_1/3_plateau_expe,Expe_quantum_SSL} for $M/M_{\sat} = 1/8$, $1/4$ and $1/3$ (see, however, Refs.~\onlinecite{sebastian-2007, Levy08} for recent experimental controversies). Miyahara and Ueda\cite{SrCu2(BO3)2_plateaux_th} explained the existence of all those plateaux as a consequence of the crystallization of excited triplets, although e.g.\ the existence of the $M/M_{\sat} = 1/8$ plateau is not confirmed in a recent investigation.\cite{dorier-2008}\\
%%%%%%%%%%%%%%%%%%%%%%%%%%%%%%%%%%%%%%%
% FIGURE
%%%%%%%%%%%%%%%%%%%%%%%%%%%%%%%%%%%%%%%
Recently, rare-earth tetraborides RB$_4$ have been attracting new interest to the SSL.\cite{TbB4,TbB4_NEW,ErB4,TmB4_1,TmB4_2,gabani-2007,siemensmeyer-2007} These compounds present a large total magnetic moment $\mathcal{J} > 1$, which justifies a \emph{classical} model although magnetic anisotropies are likely to be relevant for a description of the experiments. RB$_4$ crystallize in the tetragonal structure with space group $P 4/m b m$. R ions are placed in the $(001)$ plane on a sublattice which consists of R-R dimers that are alternatively orthogonal along the $[110]$ axis. This sublattice is again topologically equivalent to the SSL (see Fig.~\ref{fig:SSL} right panel).\\
Magnetization plateaux were found for various rational values of the saturated magnetization. Due to its large total magnetic moment ($\mathcal{J}=6$) TbB$_4$ can be considered as a classical system. Several magnetization plateaux were found, including the magnetization values $M/M_{\sat}=1/3$ and $1/2$.\cite{TbB4,TbB4_NEW} Ref.~\onlinecite{ErB4} shows that ErB$_4$ (with total magnetic moment $\mathcal{J}=15/2$) exhibits a magnetization plateau at $M/M_{\sat}=1/2$ for both $\mathbf{h}/\!/[001]$ and $\mathbf{h}/\!/[100]$. The compound TmB$_4$ also presents several magnetization plateaux for $\mathbf{h} /\!/[001]$.\cite{TmB4_1,TmB4_2,gabani-2007,siemensmeyer-2007}\\
In this paper we investigate classical Heisenberg spins on the SSL under a magnetic field. Note that we are not aware of any previous study of the classical limit in the presence of a magnetic field, although some of the first studies of the zero-field properties of the SSL did start from this limit.\cite{SS,Albrecht} In Section \ref{sec:model} we discuss the classical ground state at zero temperature and then study the role of thermal fluctuations as a scenario for the appearance of a $M/M_{\sat}=1/3$ plateau. Note that this plateau arises only for non-zero temperature and is not related to an energy-driven mechanism. To emphasize this difference and the fact that this plateau is not completly flat, we refer to it as a \emph{pseudo-plateau} in the following. Then in Section \ref{sec:Monte_Carlo} we focus on the special ratio $J'/J=1/2$ and study by means of classical Monte-Carlo simulations the influence of the system size and of the temperature on those pseudo-plateaux. A phase diagram is obtained and we draw a parallel between the SSL and the classical triangular lattice. 
Finally, in Section \ref{sec:other_Ratio} we show that a small modification $\epsilon$ of the ratio of the magnetic couplings $J'/J=1/2 + \epsilon$ slightly changes the phase diagram but still allows a pseudo-plateau to exist at $M/M_{\sat}=1/3$.

%%%%%%%%%%%%%%%%%%%%%%%%%%%%%%%%%%%%%%%%%%%%%%%%%%%%%%%%%%%%%%%%%%%%%%%%%%%%%%%%
%%%%%%%%%%%%%%%%%%%%%%%%%%%%%%%%%%%%%%%%%%%%%%%%%%%%%%%%%%%%%%%%%%%%%%%%%%%%%%%%
\section{\label{sec:model}The model}

We study the SSL in the classical limit. Heisenberg spins are represented by simple vectors $\si$ with $|\!|\si|\!|=1$. The magnetic field is applied to the $z-$axis, $\mathbf{h}=h \mathbf{e}_z$ (i.e.\ $\mathbf{h}/\!/[001]$ in the experiments carried out on RB$_4$).
The sums are taken over nearest neighbor pairs according to the edges of the squares with magnetic coupling $J'$ and over the diagonals with diagonal couplings $J$. The Hamiltonian is given by:
\be
 \mathcal{H} =  \frac{J'}{J} \sum^N_{\rm edges}\!\! \si \cdot \sj
              + \!\!\sum^N_{\rm diagonal}\!\!\!\!\! \si \cdot \sj
              - \frac{h}{J} \sum^N_{i} S_i^z \, .
\label{Hamiltonian}
\ee
By minimizing the classical energy in the absence of applied magnetic field the lowest energy configuration is found to be coplanar.\cite{SS} It is N\'eel ordered for $J'/J \ge 1$, else it is a spiral state with an angle $\varphi = \pi \pm \arccos\big(\frac{J'}{J}\big)$ between nearest-neighbor spins. Those two possible optimum values for $\varphi$ give a discrete chiral degeneracy to each triangle. The choice of the angles $\varphi$ in two neighboring triangles determines the direction of the helix. Four helices are possible which creates a supplementary four-fold discrete degeneracy in addition to the continuous one.\cite{SS} 

%%%%%%%%%%%%%%%%%%%%%%%%%%%%%%%%%%%%%%%%%%%%%%%%%%%%%%%%%%%%%%%%%%%%%%%%%%%%%%%%
\subsection{\label{subsec:GS} Classical ground state at $T=0$}
%%%%%%%%%%%%%%%%%%%%%%%%%%%%%%%%%%%%%%%
% FIGURE
\begin{figure}
 \vspace{2mm}
 \begin{tabular}{ccc}
  \includegraphics[width=0.23\textwidth]{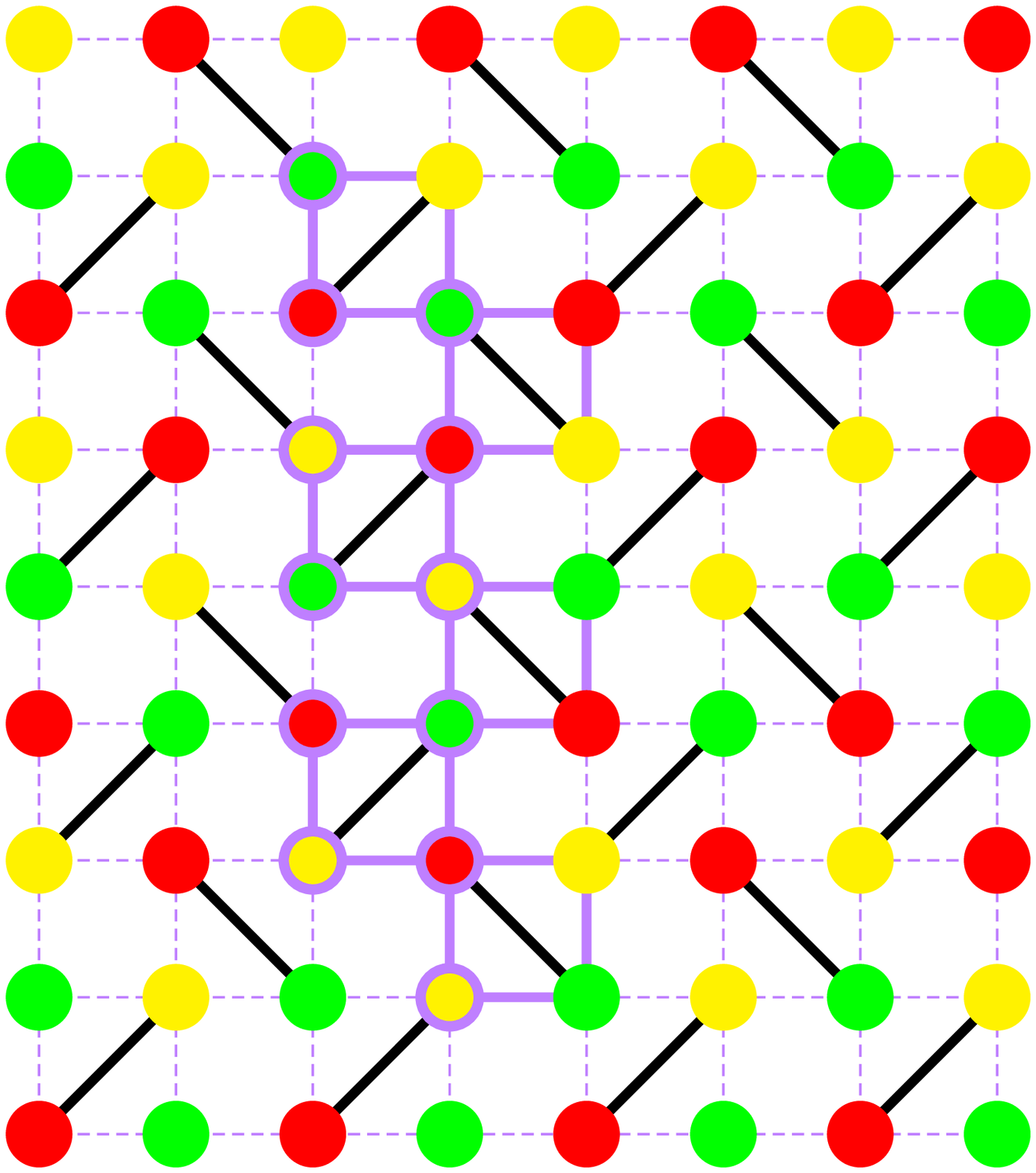} & & 
  \includegraphics[width=0.23\textwidth]{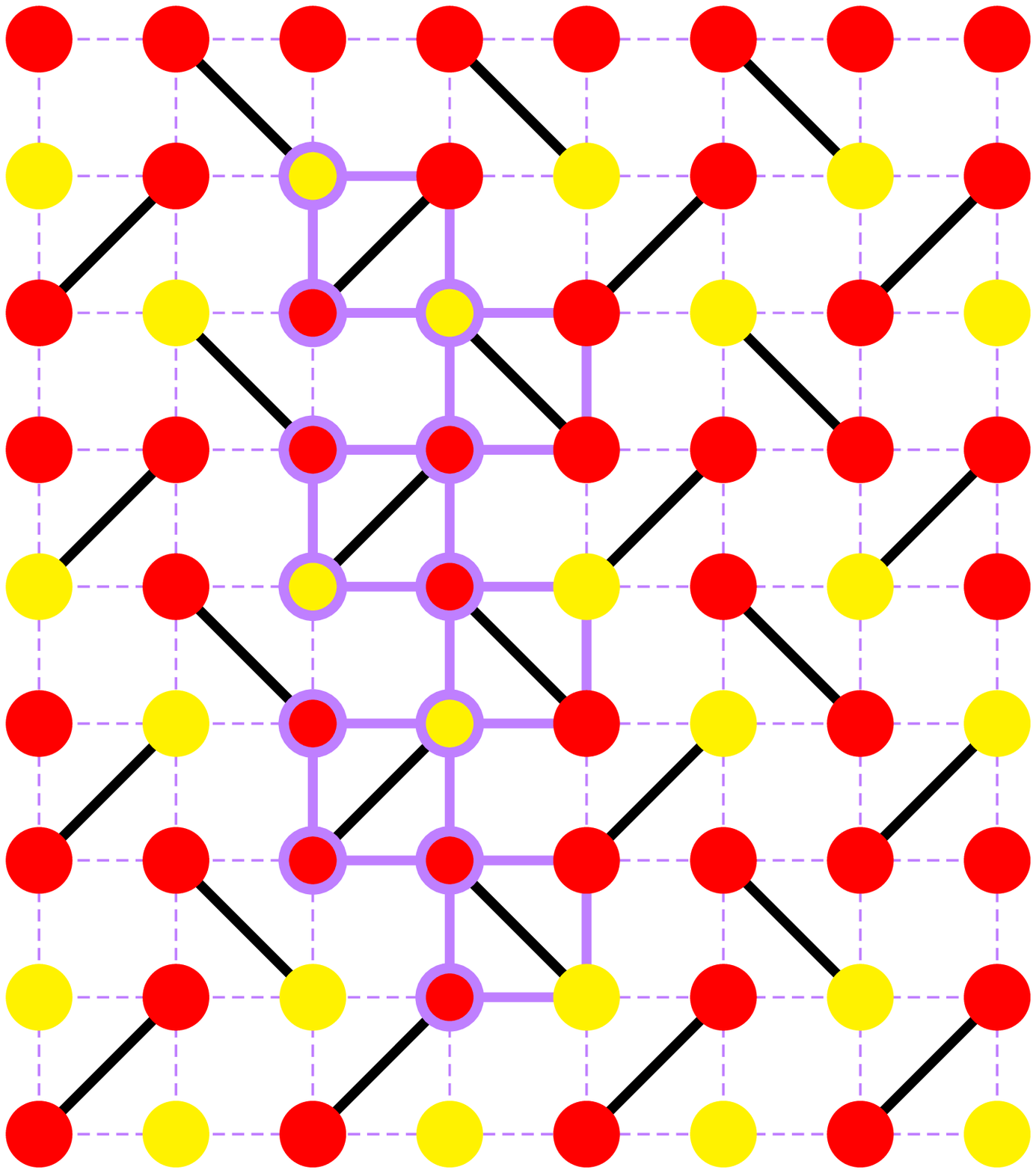} 
 \end{tabular}
 \caption{\label{fig:GS}(color online) 
Lowest energy configuration of the SSL in the $(001)$ plane for the particular ratio $J'/J=1/2$. The 12 spins unit cell is represented in bold clear lines.
Left panel: The $120\,^{\circ}$ structure with three kinds of spins orientation.
Right panel: The $UUD$ state at $M/M_{\sat}=1/3$. Each triangle contains two spins up and one spin down. Up spins are represented in red (dark gray) and down spins in yellow (light gray).}
\end{figure}
%%%%%%%%%%%%%%%%%%%%%%%%%%%%%%%%%%%%%%%
A particular ratio of the magnetic couplings allows the Hamiltonian to be expressed as a sum of elementary plaquettes.\cite{frus_sqlat_h, Benjamin} In the SSL these are triangles which share edges along the diagonal dimer bonds and otherwise corners. When the diagonal couplings are twice as big as the edge couplings (i.e.\ $J'/J=1/2$) the Hamiltonian Eq.\ (\ref{Hamiltonian}) can be written as a sum over triangles up to a constant term: 
\bea
  \mathcal{H}_{\Delta} = \frac{1}{J}\sum_{\Delta}^{N_{\Delta}}\left( \frac{J'}{2} \mathbf{S}_{\Delta}^2 - \frac{h}{3}
  \mathbf{S}_{\Delta} \right)
  \ , \ (J'/J = 1/2) \, ,
 \label{Hamiltonian_triangles}
\eea
where $\mathbf{S}_{\Delta}=\sum_{i\in \Delta} \si$ is the total spin of one triangle  and $N_{\Delta}$ is the number of triangles ($N_{\Delta}=N$). The classical ground state at zero temperature and in the absence of a magnetic field is a coplanar configuration with an angle $\varphi = \pm 2\pi/3$ between two nearest-neighbor spins. There are three possible spin orientations $\Sa$, $\Sb$, and $\Sc$ and the unit cell contains 12 sublattices (see Fig.~\ref{fig:GS} left panel). 
Minimizing the energy on a single triangle, one obtains the constraint:
\bea
  \mathbf{S}_{\Delta} = \frac{\mathbf{h}}{3J'} \ , \ (J'/J = 1/2) \, .
 \label{plaquette_constraint}
\eea
The classical ground state is obtained when this constraint is satisfied in every triangle. The saturation field $h_{\sat}$ is determined by the condition $S^z_{\Delta} = 3$ which gives $h_{\sat}=9J'$. At this value all the spins are aligned with the $z-$axis (``$UUU$ state''). We focus on the field range $0 \le h \le h_{\sat}$. According to the classical constraint Eq.\ (\ref{plaquette_constraint}) the classical ground state has only the typical global rotation as a degeneracy.\\ 
%%%%%%%%%%%%%%%%%%%%%%%%%%%%%%%%%%%%%%%
% FIGURE
\begin{figure}
 \vspace{2mm}
 \begin{tabular}{ccccccc}
  \includegraphics[height=2.5cm]{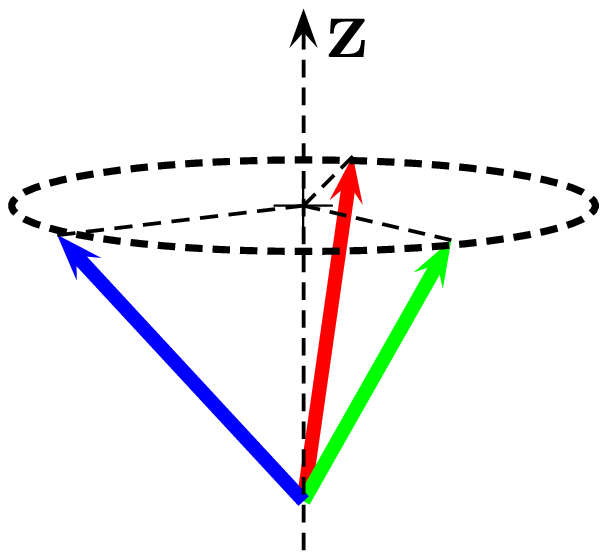} & & 
  \includegraphics[height=2.5cm]{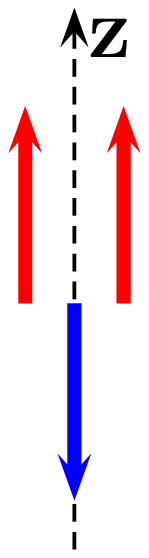} & &
  \includegraphics[height=2.5cm]{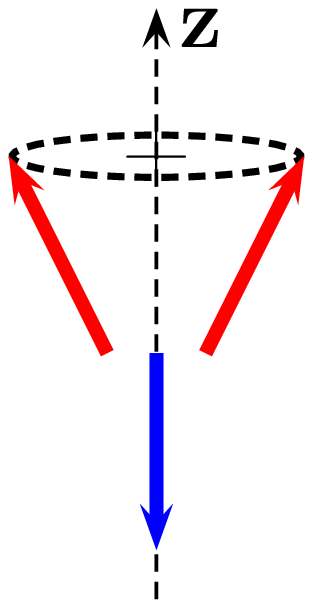} & &
  \includegraphics[height=2.5cm]{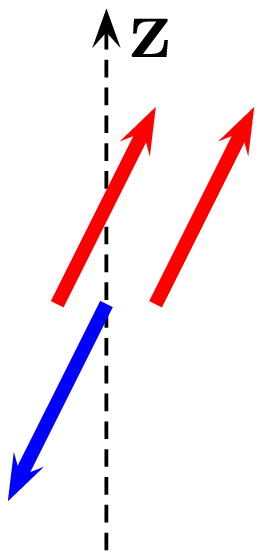} 
 \end{tabular}
 \caption{\label{fig:spin_config}(color online) From left to right: the ``umbrella'' and the ``Up-Up-Down'' states are possible spin configurations at $M/M_{\sat}=1/3$.
 The ``$Y-$state'' and the canted state are the quasi-long-range-ordered configurations discussed in Sec. \ref{subsec:R05_phase_diag}.}
\end{figure}
%%%%%%%%%%%%%%%%%%%%%%%%%%%%%%%%%%%%%%%
Let us have a closer look at the particular value $M/M_{\sat}=1/3$. The magnetic field to apply is:
\bea 
  h_{1/3} = 3J' \ , \ (J'/J = 1/2)  \, .
 \label{h_one_third}  
\eea
Stable configurations at $h=h_{1/3}$ must verify $\mathbf{S}_{\Delta} = \mathbf{e}_z$. Very different spin configurations satisfy this requirement:
the ``umbrella configuration'' and the ``Up-Up-Down'' state (see Fig.\ \ref{fig:spin_config}). In the umbrella configuration the three kinds of spins
raise as the field increases and they always have the same projection on the $z-$axis (at $M/M_{\sat}=1/3$, $\Saz=\Sbz=\Scz=1/3$). In this case, the picture of the classical ground state with three kinds of spin orientations remains until saturation (see Fig.\ \ref{fig:GS} left panel). 
On the other hand the ``Up-Up-Down'' state ($UUD$ state) is a collinear state in which each triangle contains two spins Up and one spin Down (see right panel of Fig.\ \ref{fig:GS}). This classical ground state is required in order to have a classical plateau at $M/M_{\sat}=1/3$.\cite{Hida_Affleck} 
One can easily show that at zero temperature and $M/M_{\sat}=1/3$, both configurations, umbrella and $UUD$, have the same classical energy. Since energetic considerations do not favor the $UUD$ state its existence is restricted to the field value $h_{1/3}$. Hence no magnetization plateau can appear at zero temperature.\\
Minimizing the energy on triangles naturally leads to a comparison with the classical triangular lattice, with magnetic coupling $J_{\Delta}$ between
nearest neighbors. This lattice was the subject of many theoretical studies.\cite{Lee,SP_triang_latt,triangular_latt} Kawamura and Miyashita\cite{triangular_latt} studied classical Heisenberg spins on the triangular lattice in the presence of a magnetic field along the $z-$axis. They obtained classical constraints that are strictly equivalent to Eq.\ (\ref{plaquette_constraint}) for the SSL. Minimizing the energy of a single triangle they showed that the classical ground state is completely specified by the conditions:
\bea
  |\!|\Sa|\!| = |\!|\Sb|\!| = |\!|\Sc|\!| &=& 1 \, ,\nonumber \\
  \Sa + \Sb + \Sc &=& \frac{\mathbf{h}}{3J_{\Delta}}  \ , \ (h < h_{\sat}) \, .
 \label{GS_triangular_latt} 
\eea
These constraints give information on the magnetization process of the classical triangular lattice and SSL at zero temperature. The application of the magnetic field does not break the $U(1)$ symmetry of the  $120\,^{\circ}$ configuration which means that the continuous degeneracy of the classical ground state remains. In other terms, at zero temperature and under a magnetic field, the $120\,^{\circ}$ structure raises in the umbrella configuration that closes as the field increases and reaches the $UUU$ configuration at saturation.\\
For non-zero temperature the scenario of the magnetization process is completely different. Kawamura\cite{SP_triang_latt} showed that the non-trivial degeneracy no longer persists on the triangular lattice. Even in the low-temperature limit, the most favorable spin configuration is determined not only by its energy but also by the density of states just above the ground state. This entropic selection is responsible for the appearance of new phases which will be detailed for the SSL in Section \ref{subsec:Temperature}.
%%%%%%%%%%%%%%%%%%%%%%%%%%%%%%%%%%%%%%%%%%%%%%%%%%%%%%%%%%%%%%%%%%%%%%%%%%%%%%%%
\subsection{\label{subsec:Temperature}Effect of thermal fluctuations}

The Monte Carlo (MC) simulations reported in Sec.~\ref{sec:Monte_Carlo} below indicate that, at finite temperature, the $UUD$ state is the favored configuration for a magnetic field range below and until $h=h_{1/3}$. This is possible if thermal fluctuations raise the entropy of the $UUD$ state
relatively to adjacent states. Kawamura and Miyashita\cite{triangular_latt,SP_triang_latt} showed that in the case of the triangular lattice non-zero temperature lifts the degeneracy of the classical ground state in the presence of a magnetic field. Also on the SSL only a discrete degeneracy due to the possible chiralities of the triangles remains.\\
The selection by thermal fluctuations of particular states among a degenerate manifold is called \emph{Order by disorder}.\cite{ObD,book_chapter,review_Moessner} This selection mechanism is known to be at work in various frustrated systems such as the frustrated square lattice,\cite{frus_sqlat_h} the {\kagome} lattice,\cite{ObD_Kagome, Mode_counting, Kagome_Zhitomirsky, Kagome_Shender_Holdsworth} or the checkerboard lattice.\cite{Benjamin} \\
We will see that the SSL differs in some respects from the aforementioned cases. It is nevertheless useful to recall some relevant features of the order-by-disorder selection before presenting the technical details of our analytical study of the thermal selection of the $UUD$ state. Under a small but non-zero temperature the classical spin-wave spectrum of the selected state has the particularity to present \emph{soft modes}. The energy of those modes is quartic in spin deviations instead of quadratic. 
Therefore they contribute less than ``normal'' modes to the free energy and it is clear that a state with more soft modes will be favored. More precisely, let us consider a state with classical energy $E_0$ that has $N_4$ soft modes and $N_2=N-N_4$ ``normal'' modes. The free energy is given by:
\be
 F \sim E_0 - T\left(\frac{N_2}{2} + \frac{N_4}{4}\right) \ln(T) \, .
\ee
Ref.~\onlinecite{Mode_counting} details how soft modes should be enumerated and how they affect the specific heat. In a system with no soft modes each degree of freedom brings a contribution of $k_B/2$ to the specific heat. In the quadratic approximation the contribution of the soft modes is zero.
Taking into account higher orders by adding quartic contributions each soft mode adds $k_B/4$. \\
We calculated the spectrum of the $UUD$ state in the quadratic approximation in thermal fluctuations by applying small deviations on the spin coordinates from the collinear $UUD$ state.\cite{Kagome_Shender_Holdsworth} At zero temperature the 12 sublattices of the unit cell presented in Fig.\ \ref{fig:GS} (right panel) are collinear with the $z-$axis. Under fluctuations their new coordinates are expressed in their own frame as:
\be
 \si (\R_i) = \left(\epsilon^x_i (\R_i),\epsilon^y_i (\R_i),1-\alpha_i (\R_i)
 \right) \, ,
 \label{spin_T}
\ee
where $\alpha_i=1/2\left((\epsilon^x_i)^2 + (\epsilon^y_i)^2 \right)$ is verifying the condition $|\!|\si|\!|=1$ up to quadratic order.\\
Following Ref.~\onlinecite{Mode_counting} the Hamiltonian is expanded in spin deviations from the $UUD$ state: 
\be
 \mathcal{H} = E_{UUD} + \sum_{n \ge 2} \mathcal{H}_n \, ,
\ee
where $E_{UUD}$ is the classical energy and $\mathcal{H}_n \sim \mathcal{O}(\epsilon^n)$. Up to second order in those fluctuations the Hamiltonian becomes, in Fourier space: 
\be
 \mathcal{H}_2 = \sum_{\K,v=x,y}\!\!\! \mathcal{V}_v^t(-\K) \, \mathcal{M} \, \mathcal{V}_v(\K) \, .
 \label{H2}
\ee
Fluctuations are applied to the 12 sublattices of the unit cell and the vectors $\mathcal{V}_v^t(-\K)$ ($v=x,y$) read:
\be
 \mathcal{V}_v^t(-\K) = \left(\tilde{\epsilon}^v_{1}(-\K), \dots
 ,\tilde{\epsilon}^v_{12}(-\K) \right) \, .
 \label{vector}
\ee
$\mathcal{M}$ is a $12\times12$ matrix. Fluctuations act exactly the same way on the $x$- and $y$-components and as a consequence the matrices are the same for fluctuations in the $x$- and $y$-directions. Eigenvalues should be determined by solving an order 12 polynomial which cannot be done analytically. However, we can determine the soft modes from the zeroes of the determinant which reads:
\be
\det(\mathcal{M}) \propto \left( -2 + 3\cos k_x - 3\cos 2k_x  + \cos 3k_x  +
\cos k_y  \right)^2 \, .
 \label{determinant}
\ee
This result shows that we have lines of soft modes, but no full branch. Yet this result is still exceptional since usually one finds only a finite number of soft modes.
%%%%%%%%%%%%%%%%%%%%%%%%%%%%%%%%%%%%%%%
% FIGURE
\begin{figure}
 \vspace{2mm}
 \includegraphics[width=\columnwidth]{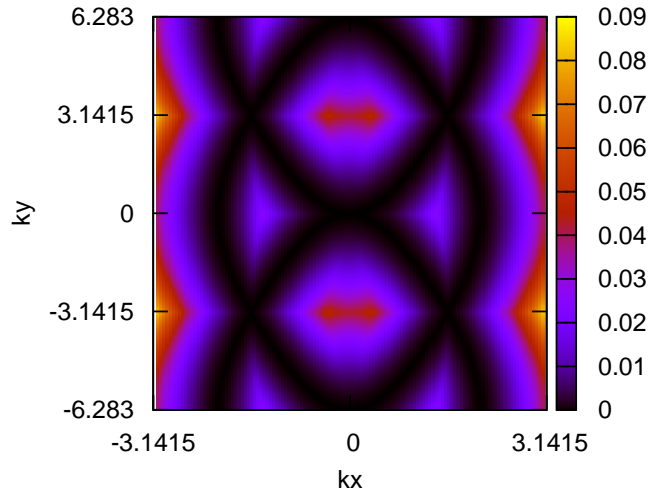} 
 \caption{\label{fig:soft}(color online) Lowest branch of the fluctuation matrix $\mathcal{M}$ above the $UUD$ state in Fourier space ($k_x$, $k_y$).
 In the darkest regions, the black lines represent the lines of soft modes whose analytical expressions are given in the text.}
\end{figure}
%%%%%%%%%%%%%%%%%%%%%%%%%%%%%%%%%%%%%%%
\\
We further performed numerical diagonalization of $\mathcal{M}$ at each point in the Fourier space $(k_x,k_y)$. For each point the smallest eigenvalue was selected in order to obtain a picture of the ground state. Fig.\ \ref{fig:soft} shows the picture obtained over a couple of Brillouin zones. The darkest regions correspond to the lowest points and we indeed observe \emph{lines} of soft modes corresponding to a cancellation of the quadratic energy.  Note that the corresponding zero eigenvalues are non-degenerate. The height of the barriers between those lines is $\Delta E \approx 0.1$ meaning that low-temperature behavior is expected to appear only for
$T \lesssim 0.1$. The analytical expression of the lines of soft modes can be obtained by finding relations between $k_x$ and $k_y$ so that the determinant
Eq.~(\ref{determinant}) vanishes:
\be
 k_x = \pm \arccos\left(\frac{1}{2} - \frac{(1\pm i\sqrt{3})(1-u(k_y))}{4u(k_y)}\right) \, ,
\ee
where $u(k_y)=(\sqrt{\cos^2 k_y - 1}-\cos k_y)^{-1/3}$.\\
Following the calculations performed by Champion and Holdsworth for the pyrochlore lattice,\cite{pyrochlore} the specific heat $C_h$ per spin is predicted to be given by:
\be
 \frac{C_h}{N k_B T} = 1 - \gamma\frac{1}{L} \, ,
\label{C_h}
\ee
where $\gamma$ is a constant related with the number of lines of soft modes.\\
The manifold of soft modes for the SSL with magnetic couplings verifying $J'/J=1/2$ scales like $L$ in a $L^2$ Fourier space. Eq.\ (\ref{C_h}) clearly shows that the drop in the specific heat, or in other terms the effect of the soft modes, will disappear in the thermodynamic limit.
\\
The lines of soft modes are reminiscent of the $\mathbf{q}=\mathbf{0}$ $UUUD$ state in the frustrated square lattice.\cite{frus_sqlat_h} However, the precise mechanism is somewhat different in the SSL since here the soft modes do not correspond to continuous deformations of the ground state. Nevertheless, the $UUD$ state is still selected for entropic reasons, like in the triangular lattice:\cite{SP_triang_latt,triangular_latt} the soft modes simply result in a free energy which is lower for the $UUD$ state than for non-collinear phases. The triangular lattice with classical Heisenberg spins does not have full branches of soft modes either, but the manifold of soft modes consists of points in Fourier space. Still, the triangular lattice exhibits pseudo-plateaux at finite temperature.
%%%%%%%%%%%%%%%%%%%%%%%%%%%%%%%%%%%%%%%%%%%%%%%%%%%%%%%%%%%%%%%%%%%%%%%%%%%%%%%%
%%%%%%%%%%%%%%%%%%%%%%%%%%%%%%%%%%%%%%%%%%%%%%%%%%%%%%%%%%%%%%%%%%%%%%%%%%%%%%%%
\section{\label{sec:Monte_Carlo}Monte Carlo simulations}
%%%%%%%%%%%%%%%%%%%%%%%%%%%%%%%%%%%%%%%
 % FIGURE
\begin{figure}
 \vspace{2mm}
 \includegraphics[width=\columnwidth]{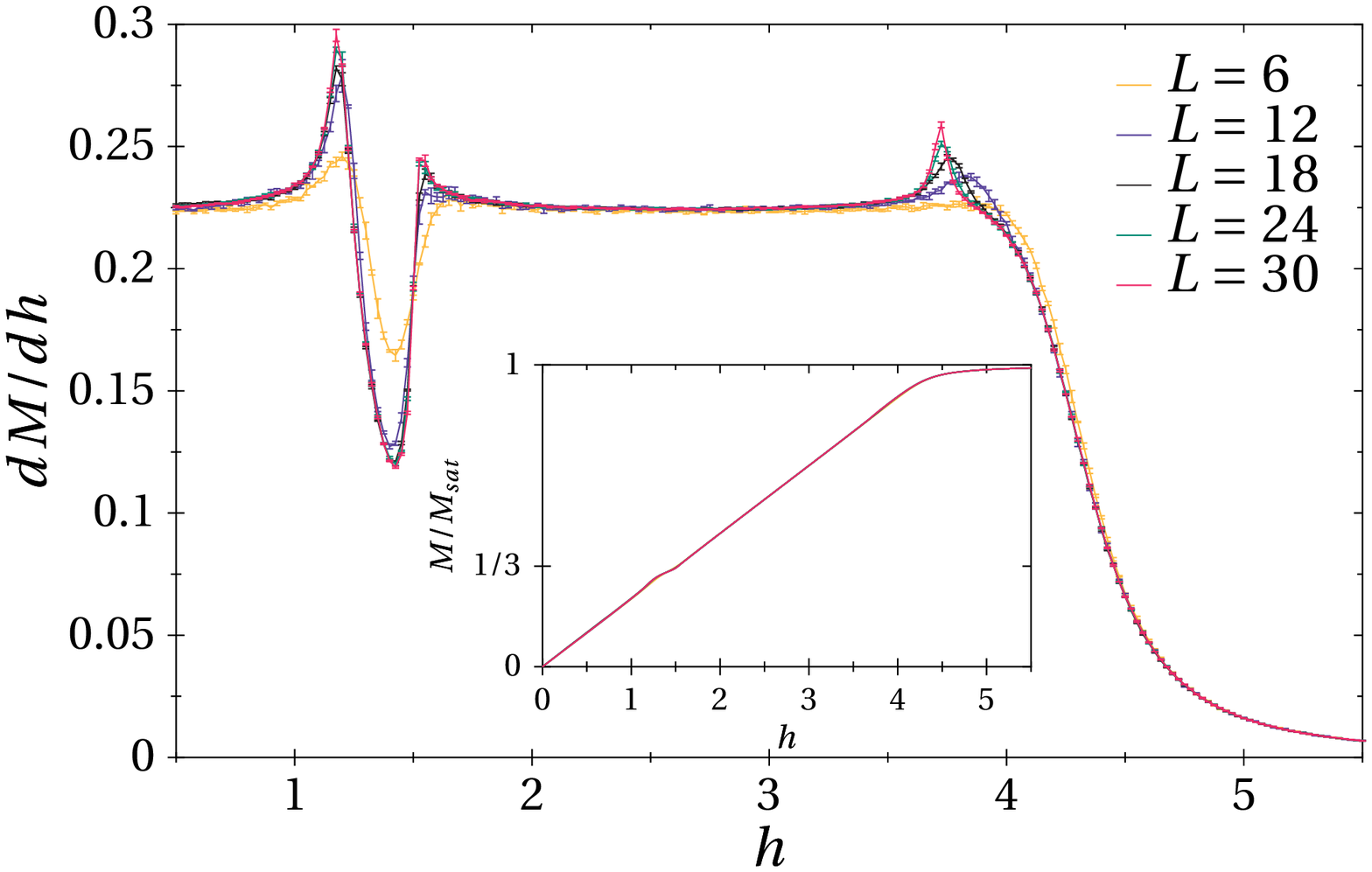}
 \caption{\label{fig:R05_size_effect}(color online) Susceptibility and magnetization (inset) of the SSL at $J'/J=1/2$ and $T=0.02$ for various system sizes.}
\end{figure}
%%%%%%%%%%%%%%%%%%%%%%%%%%%%%%%%%%%%%%%
We performed MC simulations using a standard single-spin flip Metropolis algorithm.\cite{LandauBinder} As a small refinement, we propose changes of the spin projection along and perpendicular to the field direction independently. Pseudorandom numbers were generated by the Mersenne-Twister random number generator.\cite{Mersenne} We studied square samples ($L_x=L_y=L = \sqrt{N}$) with periodic boundary conditions. Their sizes are chosen so as to be commensurate with the 12 sublattice unit cell represented in Fig.~\ref{fig:GS}: $L=6,12,18,24$, and $30$. In the simulations the diagonal coupling $J$ is set to 1. MC simulations computed in particular the magnetization, the susceptibility $\chi = dM/dh$ and the specific heat $C_h$. All physical quantities are normalized per spin. Data was collected with up to $\sim 10^{7}$ MC sweeps per point depending on the size and temperature considered. In order to insure thermal equilibrium, a comparable amount of sweeps was discarded before collecting data. Both field and temperature scans were run. Error bars are determined from independent MC runs (the average number of simulations per point is five).

%%%%%%%%%%%%%%%%%%%%%%%%%%%%%%%%%%%%%%%%%%%%%%%%%%%%%%%%%%%%%%%%%%%%%%%%%%%%%%%%
\subsection{\label{subsec:R05_curves}Magnetization pseudo-plateau at $M/M_{\sat}=1/3$}

Fig.\ \ref{fig:R05_size_effect} shows the susceptibility and the magnetization as a function of the system size at $J'/J=1/2$ for $T=0.02$. A pseudo-plateau is observed in the magnetization curve at $M/M_{\sat}=1/3$. Following the previous analytical discussion this means that the $UUD$ configuration is thermally selected at this temperature. Its existence is no longer strictly limited to one single point at $h=h_{1/3}$, but it exists for a range of magnetic field values below and until $h_{1/3}$. 
The width of those pseudo-plateaux increases with system size. One clearly observes three peaks in the susceptibility curves which become sharper and higher as the system size increases. The two first ones indicate the presence of phase transitions to the $UUD$ state while the last one corresponds to the transition to saturation.
%%%%%%%%%%%%%%%%%%%%%%%%%%%%%%%%%%%%%%%
% FIGURE
\begin{figure}
 \vspace{2mm}
 \includegraphics[width=\columnwidth]{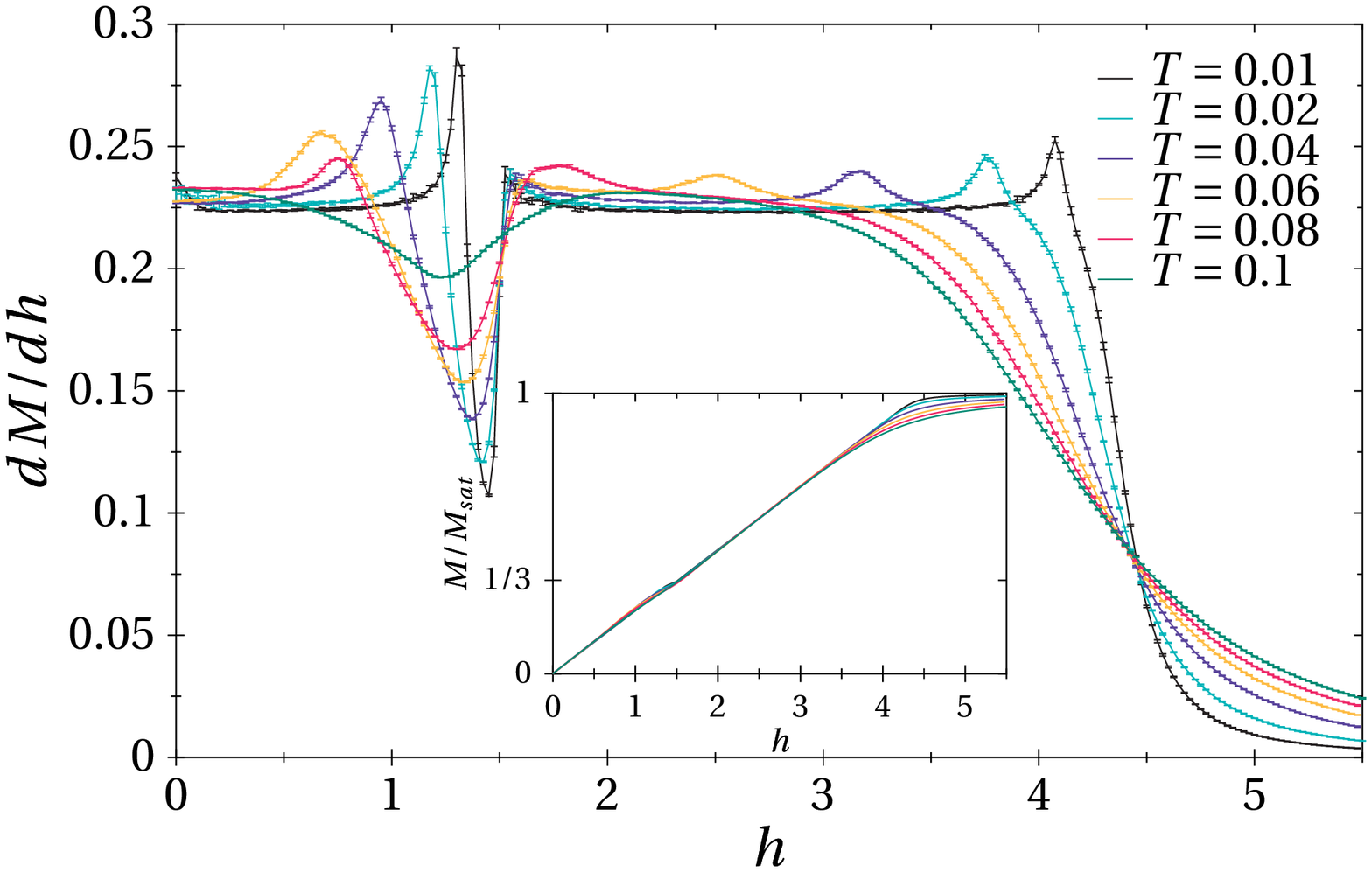}
 \caption{\label{fig:R05_T_effect}(color online) Susceptibility and magnetization (inset) of the SSL at $J'/J=1/2$ and for a system size $L=18$. The temperature $T$ varies.
}
\end{figure}
%%%%%%%%%%%%%%%%%%%%%%%%%%%%%%%%%%%%%%%
\\  
The stability of the susceptibility peak positions and height for $L \ge 18$ show that $L=18$ captures reasonably well the behavior of the plateau region in the thermodynamic limit. Fig.\ \ref{fig:R05_T_effect} shows the effect of the temperature on the susceptibility and magnetization for a $L=18$ system. The magnetization curve should tend to a straight line as the temperature decreases.
This is confirmed by the MC simulations: at very low temperature (see $T= 0.01$ in Fig.\ \ref{fig:R05_T_effect}) the pseudo-plateau is extremely small. The susceptibility presents two very sharp peaks around $h_{1/3}$. As the temperature increases the pseudo-plateau becomes broader, which confirms the entropic selection of the $UUD$ state
in the plateau phase. Finally higher temperature (see $T= 0.1$ in Fig.\ \ref{fig:R05_T_effect}) destroys it. The peaks in the susceptibility curve become rounded and then completely disappear.\\
%%%%%%%%%%%%%%%%%%%%%%%%%%%%%%%%%%%%%%%
% FIGURE
\begin{figure}
 \vspace{2mm}
 \includegraphics[width=\columnwidth]{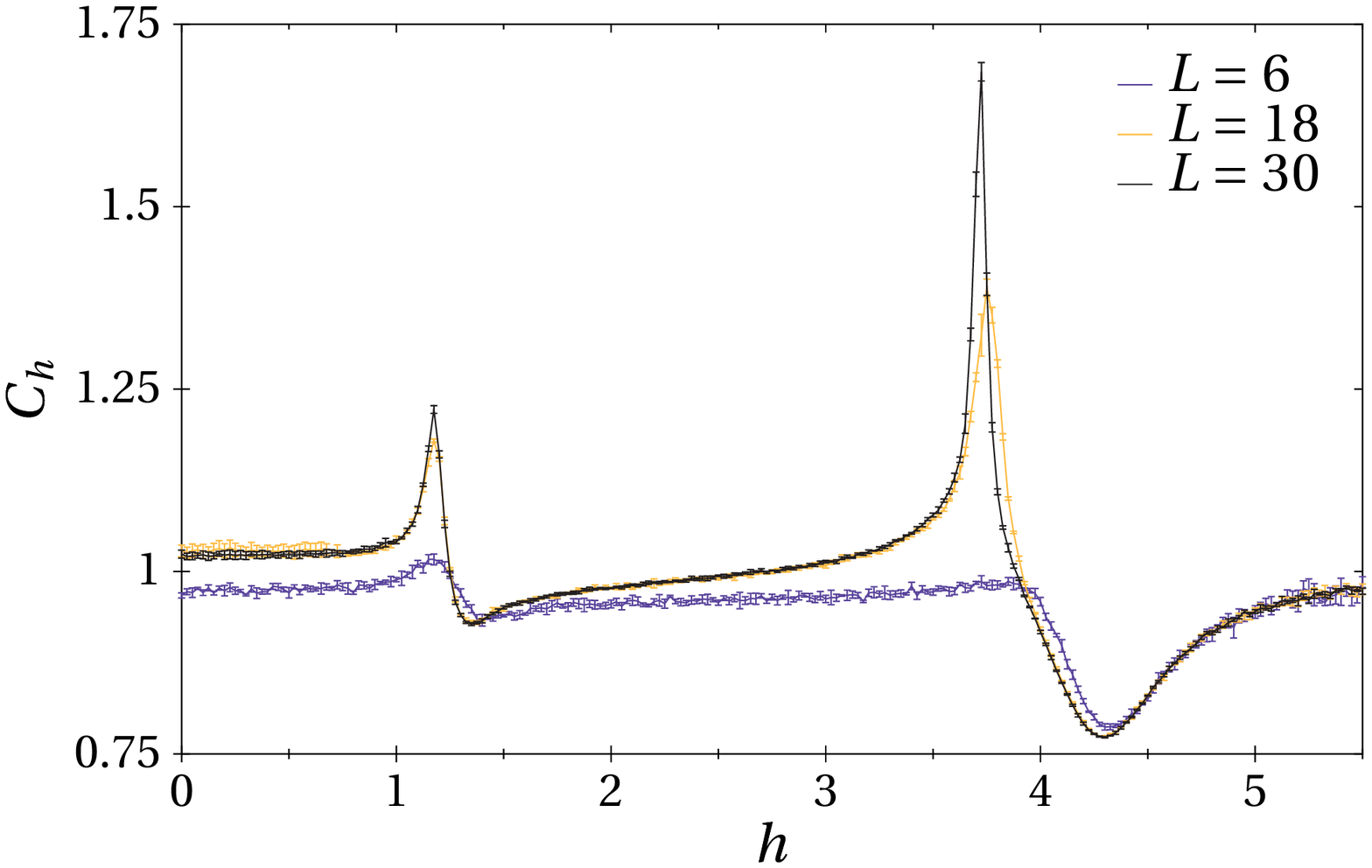}
 \caption{\label{fig:R05_Ch_size}(color online) Specific heat of the SSL with $J'/J=1/2$ at $T=0.02$ and for various system sizes.}
\end{figure}
%%%%%%%%%%%%%%%%%%%%%%%%%%%%%%%%%%%%%%%
The specific heat $C_h$ (Fig.\ \ref{fig:R05_Ch_size}) exhibits a first peak attributed to the transition from the collinear $UUD$ state to a low magnetic field state that will be detailed in Section~\ref{subsec:R05_phase_diag}. MC simulations on system sizes $L=6-42$ showed that the height of this peak scales with increasing system size as $ \sim L^{\alpha/\nu}$ with $\alpha/\nu \ll 1$ which suggests a continuous phase transition.Note that the MC simulations presented in Fig. \ \ref{fig:R05_Ch_size} were performed at $T=0.02$ which is a too high temperature to have a convergence of the specific heat towards $1$ as the system size increases, as one would expect from the equipartition theorem and Eq.~\ref{C_h}. The second peak corresponds to the transition to saturation.

%%%%%%%%%%%%%%%%%%%%%%%%%%%%%%%%%%%%%%%%%%%%%%%%%%%%%%%%%%%%%%%%%%%%%%%%%%%%%%%%
\subsection{\label{subsec:R05_phase_diag}Phase diagram for the ratio $J'/J=0.5$}

\begin{figure}
 \vspace{2mm}
 \includegraphics[width=\columnwidth]{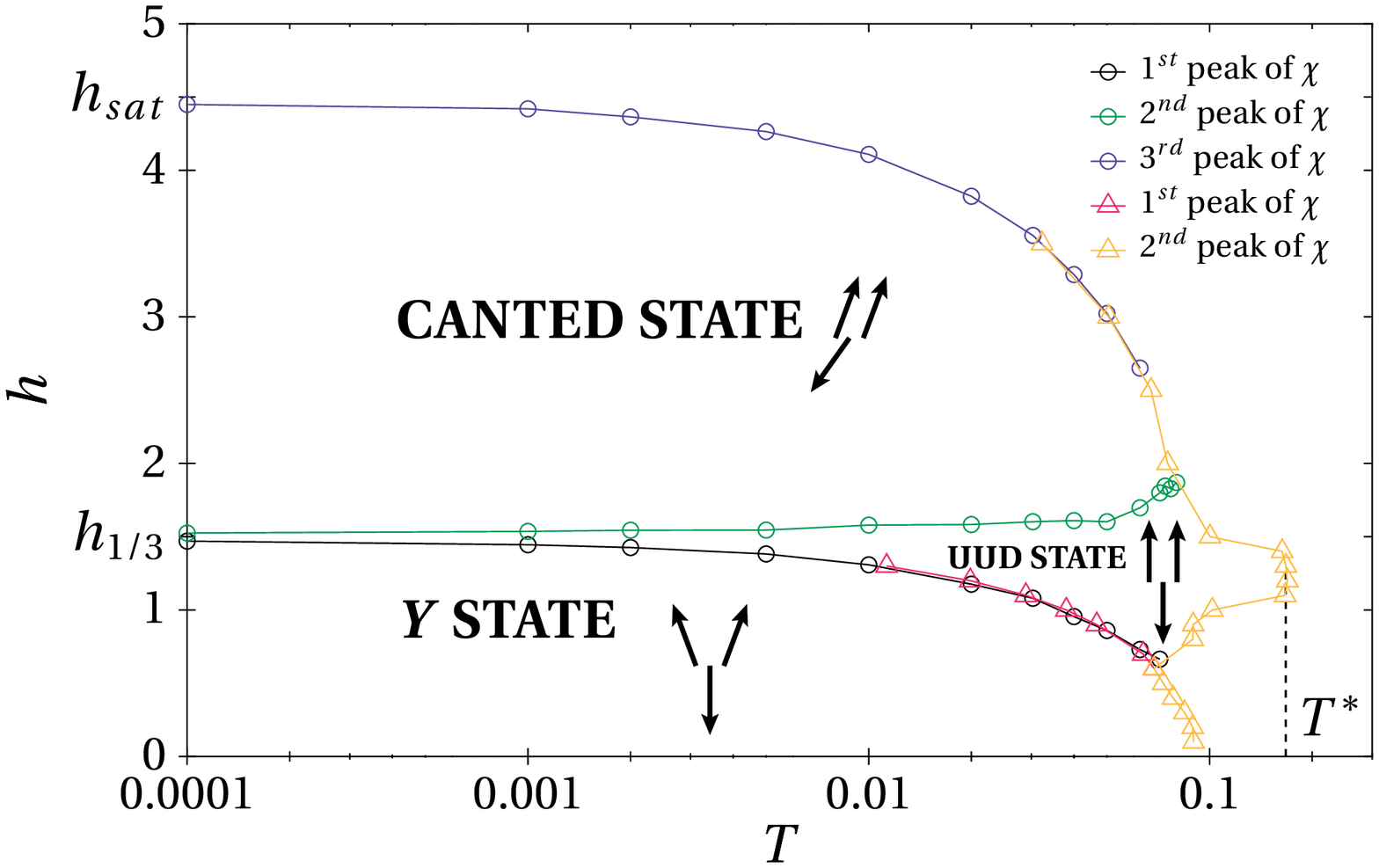}
 \caption{\label{fig:Phase_diag_R05} Phase diagram of the classical SSL for the ratio of the magnetic couplings $J'/J=1/2$.}
\end{figure}

\begin{figure}
 \vspace{2mm}
 \begin{tabular}{c}
  \includegraphics[width=0.8\columnwidth]{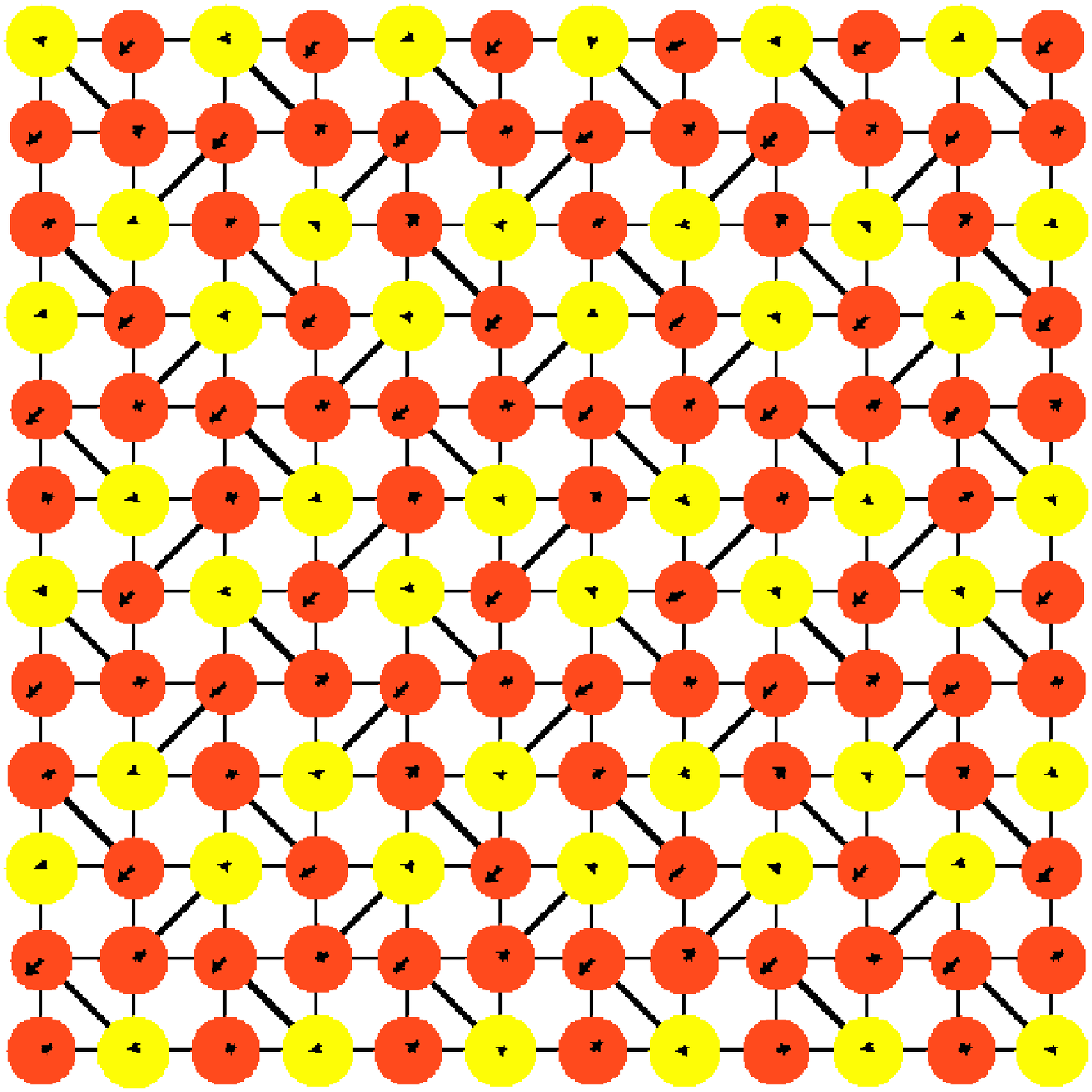} \\
  \includegraphics[width=0.8\columnwidth]{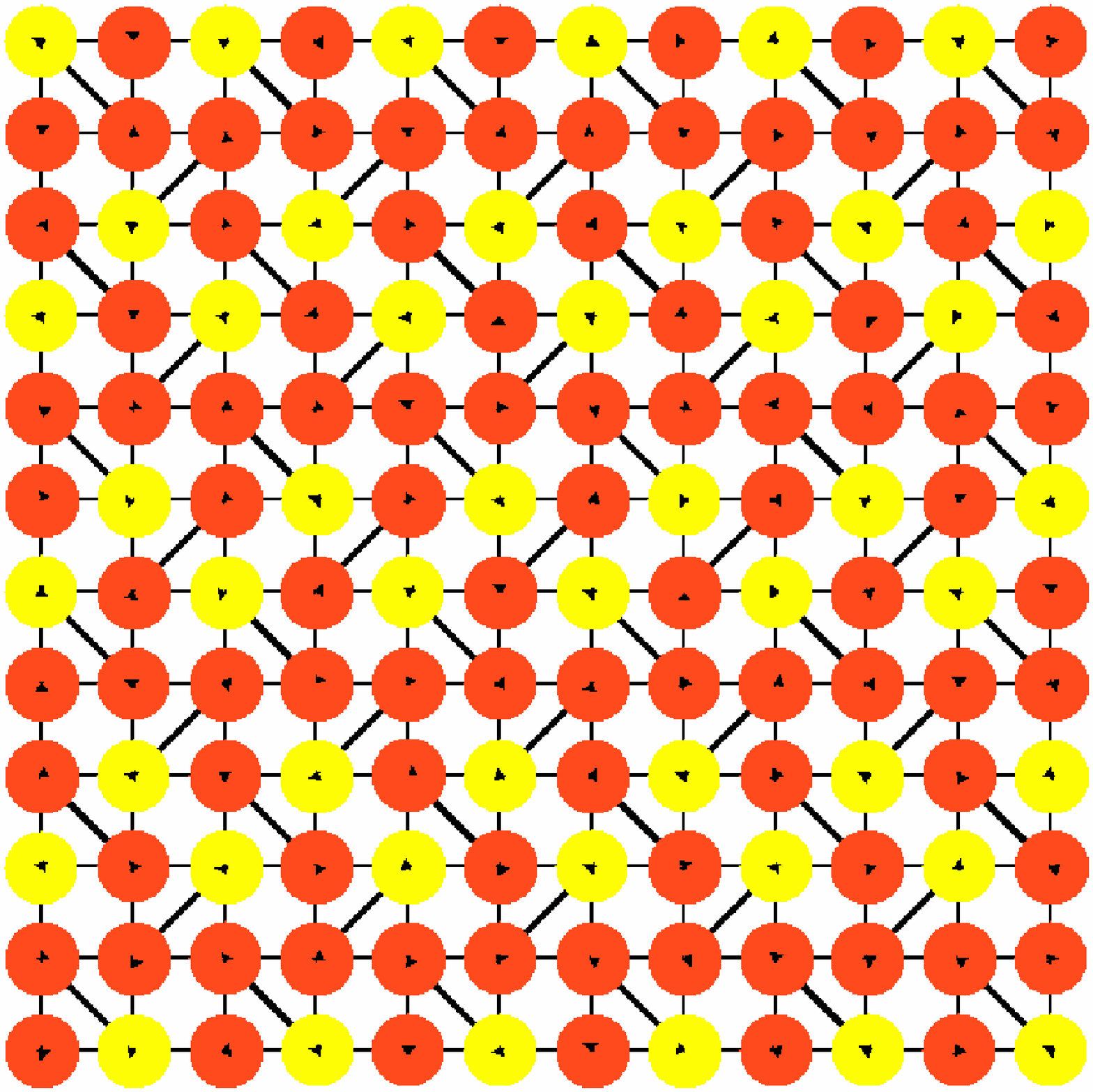} \\
  \includegraphics[width=0.8\columnwidth]{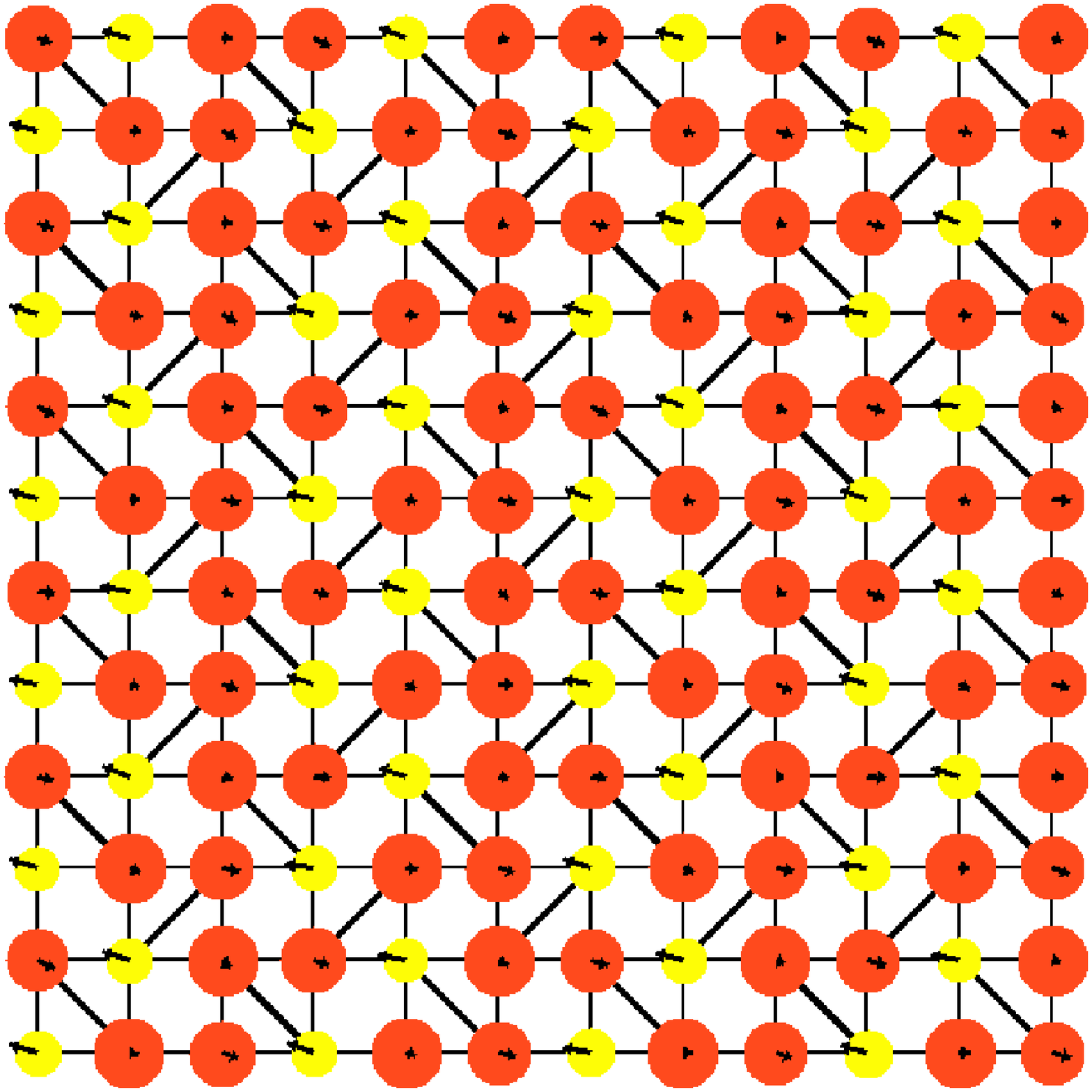} 
 \end{tabular}
 \caption{\label{fig:snapshots}(color online) Snapshots of the MC simulations of the spin texture in a $L=12$ system for $J'/J=1/2$ and $T \sim 0.003$. The circles represent the projection on the $z-$axis, red (dark gray) if $S^z>0$, yellow (light gray) otherwise. The radius is proportional to $|S^z|$. The arrow denote the projection in the lattice plane. From upper to lower panel: $Y$-configuration at $h=1.2$, $UUD$ configuration at $h_{1/3}=1.5=h_{\rm sat}/3$ and canted state at $h=1.8$.}
\end{figure}

Next we discuss the domain of existence of the $UUD$ collinear phase in the $(h,T)$ plane. Up to 5 MC simulations per point were performed on a $L=12$ ($N=144$) system which gives a good qualitative picture of the position and nature of the phases. Both temperature and magnetic field scans were performed. The boundaries of the phases are determined by the positions of the peaks in the susceptibility $\chi$. Fig.\ \ref{fig:Phase_diag_R05} shows this schematic phase diagram in the plane $(h, T)$.
Numbers are attributed to the three susceptibility peaks observed from low to high magnetic field in magnetic field scans of Fig. \ref{fig:R05_T_effect} (curves with circles) and to the two peaks from low to high temperature in temperature scans (curves with triangles).\\
We simulated by means of MC the spin texture for $L=12, 18$ systems in order to obtain a qualitative picture of the spin configuration in the phases below and above the $UUD$ phase. Fig.~\ref{fig:snapshots} shows some snapshots of the spin configurations in a $L=12$ system at $T \sim 0.003$. Like in the classical triangular lattice the magnetization process at finite temperature is very different from the one at zero temperature. The phase in the low-field region is the ``$Y$ configuration'' (see Fig.~\ref{fig:spin_config}). In this configuration each triangle contains two spins above the $(x,y)$ plane (red dark circles in Fig.~\ref{fig:snapshots}) while the last one is pointing down and is almost collinear with the $z-$axis (yellow light circles in Fig.~\ref{fig:snapshots}). This state is characterized by a single angle between the two spins pointing in the positive $z-$direction. The $Y$ configuration breaks the rotational symmetry around the $z-$axis and as a consequence of the Mermin-Wagner theorem\cite{MW} cannot be long-range ordered for $T>0$. The spins above the $(x,y)$ plane are at most quasi-long-range-ordered with correlation functions that decay algebraically. On the other hand one can expect long-range order on the remaining down spin sublattices.\\
%%%%%%%%%%%%%%%%%%%%%%%%%%%%%%%%%%%%%%%
% FIGURE
%%%%%%%%%%%%%%%%%%%%%%%%%%%%%%%%%%%%%%%
Then in a magnetic field range below and until $h_{1/3}$ the collinear $UUD$ state is the lowest energy configuration (see center panel of Fig.~\ref{fig:snapshots}). As the temperature goes to zero, the width of this phase converges to a single point at exactly $h=h_{1/3}=3/2$ which is in complete agreement with the analytical prediction. This phase does not break the $U(1)$ symmetry and hence true long-range order is realized.\\
In the high-field region the ground state is the ``canted state'' (see Fig.~\ref{fig:spin_config}). In this state the spins down are raising as the
magnetic field increases while the spins up are no longer collinear with the $z-$axis. This state is characterized by two angles: one between the $z-$axis and the two spins above the $(x,y)$ plane and one between the $z-$axis and the single spin below the $(x,y)$ plane (see bottom panel of Fig.~\ref{fig:snapshots}). As in the $Y$ phase, at most quasi-long-range order is expected at finite temperature.\\
%%%%%%%%%%%%%%%%%%%%%%%%%%%%%%%%%%%%%%%
% FIGURE
%%%%%%%%%%%%%%%%%%%%%%%%%%%%%%%%%%%%%%%
In the higher field region the system reaches saturation and all spins are pointing up collinearly with the $z-$axis ($UUU$ state). The zero-temperature limit of the saturation field is $h_{\sat}=9/2$, as derived in Section~\ref{subsec:GS}. At higher temperature the system enters the disordered paramagnetic phase. The highest temperature where the $UUD$ state still exists $T^*$ is estimated as $T^* \approx 0.17$. Note that the classical spin-wave spectrum contains energy barriers between lines of soft modes of the same order of magnitude. Hence this temperature agrees with the range of temperatures in which we expected the low-temperature behavior to appear. \\
The phase diagram of the SSL with the particular coupling ratio $J'/J=1/2$ presents similarities with the one of the classical triangular\cite{triangular_latt} and {\kagome}\cite{Kagome_Zhitomirsky} lattices.  Watarai {\it et al.}\cite{entropy_hex_latt} suggested that in the classical triangular lattice with Heisenberg spins these transitions could be of the second order.  The scaling of the specific heat shown in Fig.\ \ref{fig:R05_Ch_size} is in agreement with a continuous transition. We propose that a Berezinski\u{\i}-Kosterlitz-Thouless transition\cite{Berezinskii,KT_2,KT_1} takes place as the system enters the $UUD$ phase. Following what was obtained in the case the triangular lattice, the transition from the collinear $UUD$ state to the disordered phase should belong to the universality class of the three-state Potts model.\cite{Potts_model, review_Potts}
%%%%%%%%%%%%%%%%%%%%%%%%%%%%%%%%%%%%%%%%%%%%%%%%%%%%%%%%%%%%%%%%%%%%%%%%%%%%%%%%
%%%%%%%%%%%%%%%%%%%%%%%%%%%%%%%%%%%%%%%%%%%%%%%%%%%%%%%%%%%%%%%%%%%%%%%%%%%%%%%%
\section{\label{sec:other_Ratio}Study of the ratio $J'/J=1/2 + \epsilon$}

\begin{figure}
 \vspace{2mm}
 \includegraphics[width=\columnwidth]{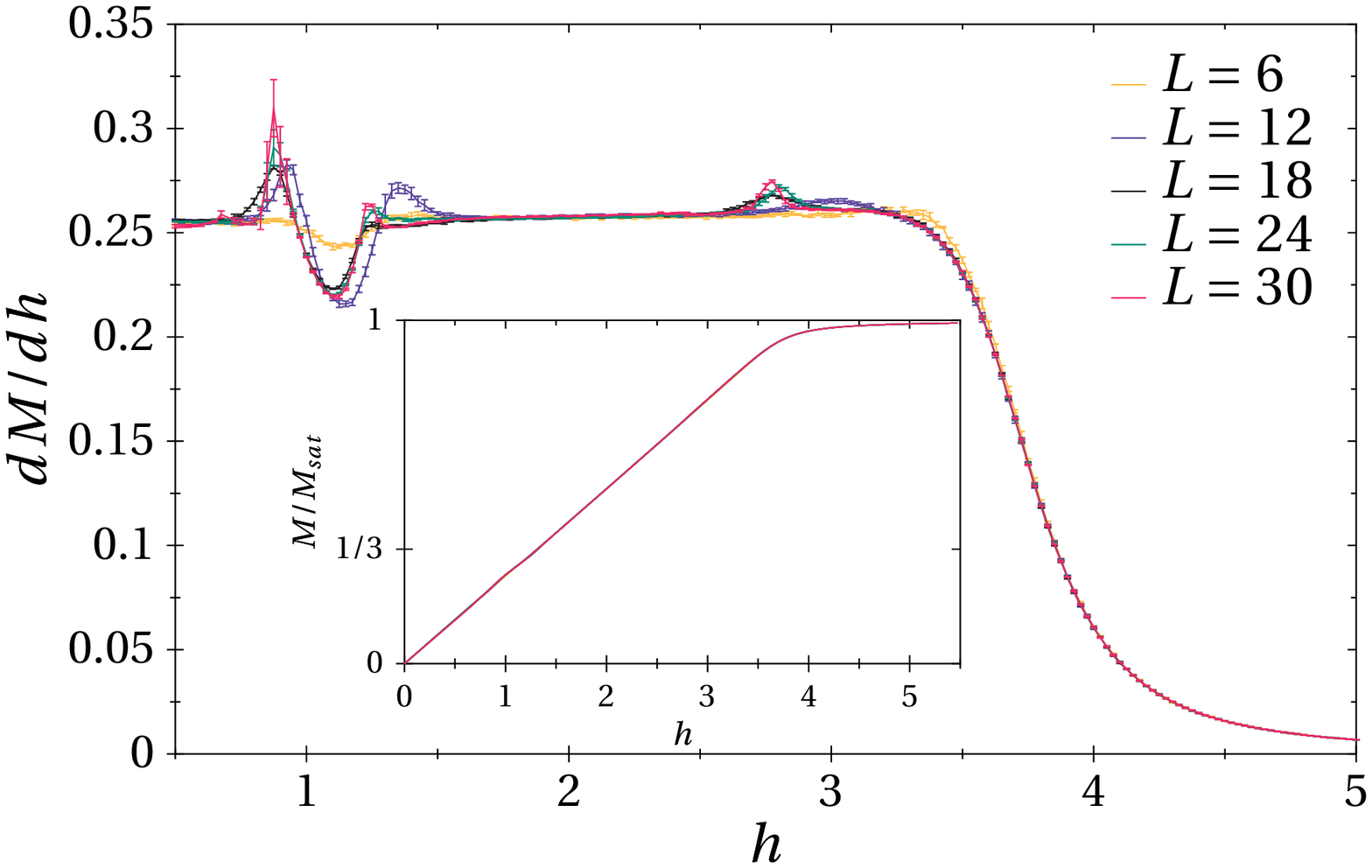}
 \caption{\label{fig:Plots_R04}(color online) MC results for the susceptibility and magnetization (inset) of the SSL at $J'/J=0.4$ and $T=0.02$ for various system sizes.}
\end{figure}

\begin{figure}
 \vspace{2mm}
 \includegraphics[width=\columnwidth]{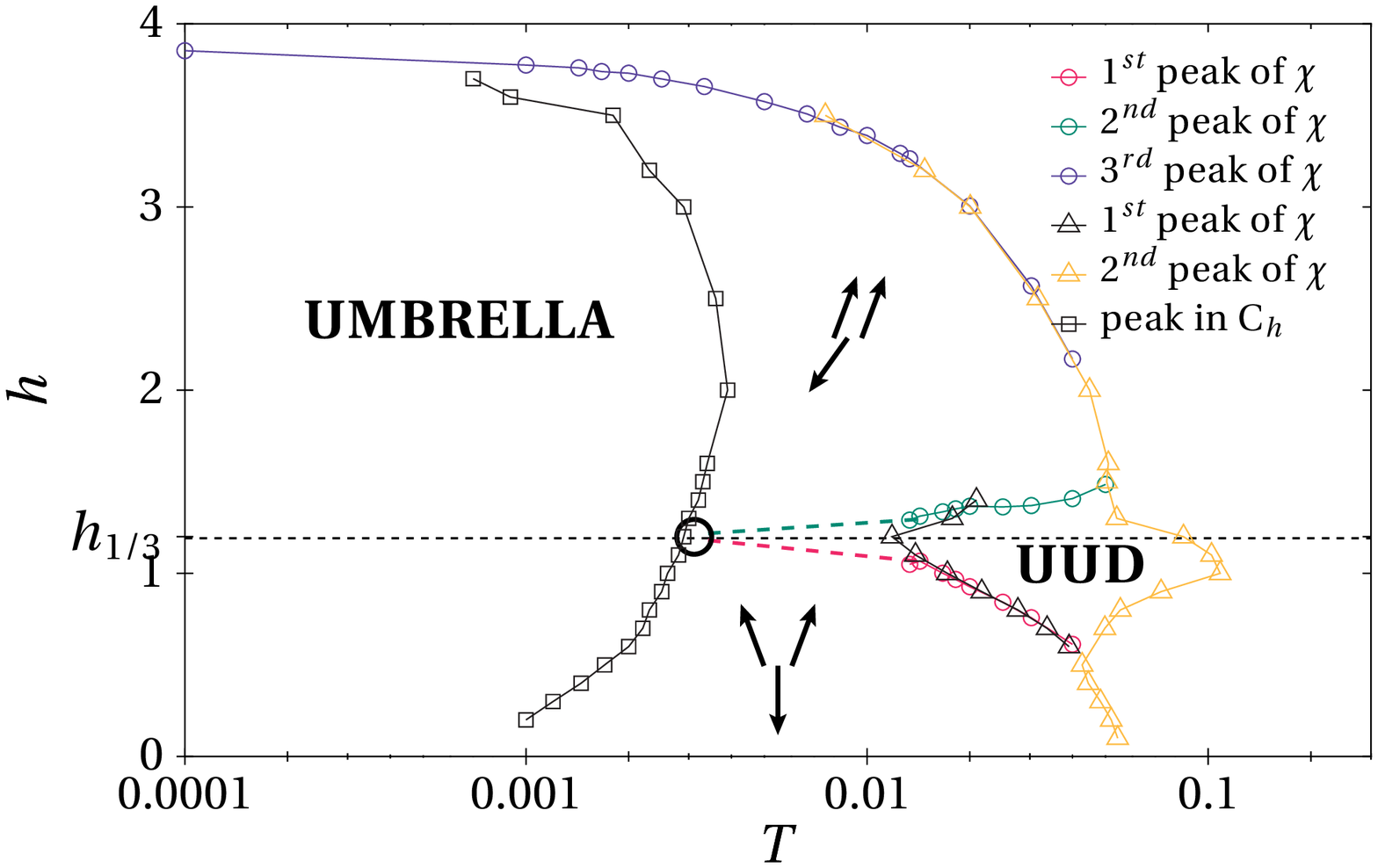}
 \caption{\label{fig:Phase_diagram_R04_L12}(color online) Phase diagram for the SSL for the ratio
 $J'/J=0.4$.}
\end{figure}

In this section we consider now a small deviation $\epsilon$ from the ratio $J'/J=1/2$. In the following discussion we use Monte Carlo results for $J'/J=0.4$ (i.e.\ $\epsilon=-0.1$). We observed that qualitatively similar behavior also appears for the other sign of $\epsilon$. \\
We performed MC simulations on systems sizes $L=6,12,18,24$, and $30$. Data were collected with up to $10^{7}$ MC steps per point. Fig.\ \ref{fig:Plots_R04} shows the susceptibility and the magnetization as a function of the magnetic field at the same temperature as in Fig.\ \ref{fig:R05_size_effect} for the ratio $J'/J=1/2$. The susceptibility still presents two peaks around $M/M_{\sat} = 1/3$ and a pseudo-plateau is observed in the magnetization curve. Therefore one can expect $UUD$ to be the selected configuration at $T=0.02$ even for ratios $J'/J=1/2 + \epsilon$.
According to the position of the peaks of the susceptibility, the width of the pseudo-plateau increases as the system size increases.\\
Following exactly the same calculations as the ones presented in Section~\ref{subsec:Temperature} for $J'/J=1/2$, we applied thermal fluctuations on top of the $UUD$ state for the ratio $J'/J=0.4$. The fluctuations matrix $\mathcal{M}$ exhibits negative eigenvalues which means that the $UUD$ configuration is no longer the favored configuration in the quadratic approximation for the spin deviations. In other words, the collinear $UUD$ state is no longer selected at ``very low temperatures'' for $J'/J\ne1/2$. 
%%%%%%%%%%%%%%%%%%%%%%%%%%%%%%%%%%%%%%%
% FIGURE
%%%%%%%%%%%%%%%%%%%%%%%%%%%%%%%%%%%%%%%
We used MC simulations in order to obtain a picture of the spin configuration in the temperature range $T<0.001$. It turns out that the favored configuration is an umbrella that is closing until saturation as the magnetic field increases. At ``higher temperature'' ($T \ge 0.01$) we observed that for $h \approx h_{1/3}$ the system enters the $UUD$ phase. This ``high temperature'' regime is not analytically accessible with lowest-order thermal fluctuations.\\ 
Following the same procedure as in Sec.\ \ref{subsec:R05_phase_diag}, we performed magnetic field and temperature scans in order to determine the region in which the $UUD$ state survives. Fig.\ \ref{fig:Phase_diagram_R04_L12} shows a schematic phase diagram for the SSL at $J'/J=1/2 + \epsilon = 0.4$ in the $(h,T)$ plane.
Five separate simulations per point were performed on $L=12$ systems. MC data were collected up to $10^7$ MC steps per point. Temperature (circles) and field (triangles) scans were performed with the same ordering of the susceptibility peaks as in Section~\ref{subsec:R05_phase_diag}. As one expects from the analytical arguments, the resulting phase diagram differs from the case $J'/J=1/2$. In the lowest temperature region, the system undergoes the same kind of magnetization process as at zero temperature: the coplanar spiral becomes an umbrella configuration as the magnetic field increases. The angle between two nearest-neighbor spins is $\pi \pm \arccos\left(1/2 + \epsilon\right)$. Hence, the spiral at zero temperature is not exactly commensurate with a 12 spin unit cell. However, as the temperature increases the spin positions fluctuate and a 12 sublattice unit cell is recovered in average on the spin positions. Hence we suggest that the transition from the umbrella to the phases with 12 sublattices (squares in Fig.\ \ref{fig:Phase_diagram_R04_L12}) is an incommensurate-commensurate phase transition.\\
%%%%%%%%%%%%%%%%%%%%%%%%%%%%%%%%%%%%%%%
% FIGURE
%%%%%%%%%%%%%%%%%%%%%%%%%%%%%%%%%%%%%%%
At higher temperature we recover the three phases that were obtained for $J'/J=1/2$ in different magnetic field regions. In the low-field region
we still obtain the ``$Y$ configuration'' and in the high-field region, before saturation, the ``canted state''. The long-range-ordered $UUD$ collinear state still survives in a small region between the two aforementioned phases. For $T<0.01$ the $UUD$ phase is expected to become narrower as the temperature decreases (dashed bold lines). A precise determination how the four phases merge (black circle) is beyond the scope of this work.

%%%%%%%%%%%%%%%%%%%%%%%%%%%%%%%%%%%%%%%%%%%%%%%%%%%%%%%%%%%%%%%%%%%%%%%%%%%%%%%%
%%%%%%%%%%%%%%%%%%%%%%%%%%%%%%%%%%%%%%%%%%%%%%%%%%%%%%%%%%%%%%%%%%%%%%%%%%%%%%%%
\section{\label{sec:Conclusion}Conclusion}

We studied the magnetization process of the Shastry-Sutherland lattice in the classical limit. We found a pseudo-plateau at $1/3$ of the saturated magnetization which corresponds to a collinear $UUD$ state. The spectrum of spin waves above this state has lines of soft modes, like the $\mathbf{q}=\mathbf{0}$ $UUUD$ state on the frustrated square lattice.\cite{frus_sqlat_h} However, in contrast to the frustrated square lattice, the $M/M_{\rm sat} = 1/3$ ground state of the SSL has no local continuous degeneracies. Therefore, the selection mechanism of the $UUD$ state in the SSL is more similar to the triangular lattice.\cite{triangular_latt,SP_triang_latt} It may be interesting to note that also the magnetization curve of the Ising model on the Shastry-Sutherland lattice exhibits exactly one plateau with $M/M_{\sat}=1/3$.\cite{meng-2008, Ising_SSL_2008}\\
Furthermore, we performed Monte-Carlo simulations and obtained a phase diagram in the $(h,T)$ plane for the particular magnetic coupling ratio $J'/J=1/2$. We confirmed that the Shastry-Sutherland lattice behaves like the triangular lattice and it presents two quasi-long-range-ordered phase below and above the long-ranged-ordered $UUD$ phase. We also showed that the pseudo-plateau survives for a small variation $\epsilon$ of the magnetic coupling ratio around $J'/J=1/2$. The phase diagram was found to present a new incommensurate umbrella phase in the low-temperature region, at least for $J'/J=0.4$.\\
In the future, it would be interesting to compute order parameters \cite{Kagome_Zhitomirsky} and characterize the phase transitions of the SSL with classical Heisenberg spins. Furthermore, magnetic anisotropies are also expected to be important for the rare-earth tetraborides.\cite{TbB4,TbB4_NEW, ErB4,TmB4_1,TmB4_2,gabani-2007,siemensmeyer-2007} Finally, it has been observed on some other lattices \cite{Penc,lattice-chain,Wang_PRL} that coupling to phonons leads to a stabilization of magnetization plateaux. It would therefore be interesting to study the classical Heisenberg model coupled to lattice degrees of freedom also on the Shastry-Sutherland lattice.

%%%%%%%%%%%%%%%%%%%%%%%%%%%%%%%%%%%%%%%%%%%%%%%%%%%%%%%%%%%%%%%%%%%%%%%%%%%%%%%%
%%%%%%%%%%%%%%%%%%%%%%%%%%%%%%%%%%%%%%%%%%%%%%%%%%%%%%%%%%%%%%%%%%%%%%%%%%%%%%%%
\begin{acknowledgments}
D.C.C., A.H., and P.P.\ would like to thank M.~E.\ Zhitomirsky for collaboration during an early stage of this work. Furthermore, we are grateful to R.\ Moessner and H.\ Kawamura for helpful discussions. This work was partially supported by the ESF through HFM grants 1872 and 2268 as well as by the DFG through grant HO~2325/4-1.
\end{acknowledgments}
%%%%%%%%%%%%%%%%%%%%%%%%%%%%%%%%%%%%%%%%%%%%%%%%%%%%%%%%%%%%%%%%%%%%%%%%%%%%%%%%
%%%%%%%%%%%%%%%%%%%%%%%%%%%%%%%%%%%%%%%%%%%%%%%%%%%%%%%%%%%%%%%%%%%%%%%%%%%%%%%%
%\newpage %Just because of unusual number of tables stacked at end

%%%%%%%%%%%%%%%%%%%%%%%%%%%%%%%%%%%%%%%%%%%%%%%%%%%%%%%%%%%%%%%%%%%%%%%%%%%%%%%%
%%%%%%%%%%%%%%%%%%%%%%%%%%%%%%%%%%%%%%%%%%%%%%%%%%%%%%%%%%%%%%%%%%%%%%%%%%%%%%%%
%%%%%%%%%%%%%%%%%%%%%%%%%%%%%%%%%%%%%%%%%%%%%%%%%%%%%%%%%%%%%%%%%%%%%%%%%%%%%%%%
\end{document}